\documentclass{WileyMSP-template}
\usepackage{amssymb}
\usepackage{pdfpages}
\usepackage{float}
\usepackage{etoolbox}   

\newtoggle{marking}
\togglefalse{marking}    

\newcommand{\marktext}[1]{%
  \iftoggle{marking}{\marktext{#1}}{#1}%
}

\usepackage{soul} 
\usepackage{xcolor} 

\setulcolor{blue}

\begin{document}

\pagestyle{fancy}
\rhead{\includegraphics[width=2.5cm]{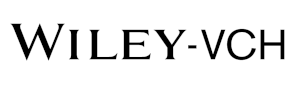}}

\title{Ultra-compact broadband spot size converter using \\ metamaterial cell-based inverse design}

\maketitle


\author{Alejandro Sánchez-Sánchez*}
\author{Carlos Pérez-Armenta}
\author{José Manuel Luque-González}
\author{Alejandro Ortega-Moñux}
\author{J.Gonzalo Wangüemert-Pérez}
\author{Íñigo Molina-Fernández}
\author{Robert Halir}

\begin{affiliations}
Alejandro Sánchez-Sánchez, Carlos Pérez-Armenta, José Manuel Luque-González, Alejandro Ortega-Moñux, J.Gonzalo Wangüemert-Pérez, Íñigo Molina-Fernández, Robert Halir\\
Telecommunication Research Institute (TELMA), Universidad de Málaga, E.T.S.I. Telecomunicación, 29010 Málaga, Spain\\ 
Email Address: as.sanchez@uma.es

\end{affiliations}



\begin{abstract}
With the expansion of silicon photonics from datacom applications into emerging fields like optical I/O, quantum and programmable photonics, there is an increasing demand for devices that combine ultra-compact footprints, low losses, and broad bandwidths. While inverse design techniques have proven very efficient in achieving small footprints, they often underutilize physical insight and rely on large parameter spaces that are challenging to explore, thereby limiting the performance of the resulting devices. Here we present a design methodology that combines inverse design with a topology based on cells, each of which contains a subwavelength metamaterial. This approach significantly reduces the parameter space, while the inherent anisotropy of the subwavelength structures yields shorter devices. We experimentally demonstrate our technique with an ultra-compact spot size converter that achieves a ×24 expansion ratio times (from 0.5\,µm  to 12\,µm) over a length of only 7.2\,µm, with insertion losses of 0.8\,dB across a measured bandwidth of 160\,\,nm (up to 300\,\,nm in simulation), surpassing the state-of-the-art by a wide margin. 
\end{abstract}


\section{Introduction}

Over the last decade, silicon-on-insulator has quickly developed into a mainstream photonics integration platform with applications spanning from high-speed optical communications to sensing and lidar imaging \cite{shekhar_roadmapping_2024,torres-cubillo_high-performance_2024,li_progress_2022}. Emerging fields such as optical I/O for high speed connectivity in AI systems, quantum  and programmable photonics \cite{wang_co-designed_2024, pelucchi_potential_2021,bogaerts_programmable_2020}, which rely on photonic integrated circuits with a large number of components, increasingly require low-loss devices with very small footprints. Two strategies, which have evolved separately over the last years, can be exploited to tackle this challenge. On the one hand, inverse design techniques which rely on purely numerical geometry optimization with little to no use of physical insight have proven extremely powerful to design very compact devices \cite{campbell_review_2019,molesky_inverse_2018,lalau-keraly_adjoint_2013}.  These approaches are usually based on dividing the design area into “pixels” (the smallest features that can be reliably fabricated), and then optimizing each of them, resulting in a large number of parameters. Inverse design therefore requires significant computational power \cite{li_intelligent_2024}, limiting the size of tractable devices. This often results in designs that exhibit limited performance in terms of insertion losses and/or bandwidth. On the other hand, subwavelength based metamaterials, offering lithographically controlled optical properties, have emerged as a universal tool to achieve devices with breakthrough performance \cite{cheben_subwavelength_2018,luque-gonzalez_review_2021,cheben_recent_2023,staude_metamaterial-inspired_2017}. Relying heavily on physical modelling and insight into their behavior, these structures have enabled the design of myriad high-performance, broad bandwidth devices, including on-chip lenses \cite{luquegonzalez_ultracompact_2019}, mode multiplexers  \cite{yi_-chip_2023} and polarization beam splitters \cite{xu_metastructured_2023}, to name a few. The inherent anisotropy of these structures furthermore results in shorter devices \cite{halir_ultrabroadband_2016}. However, their use depends entirely on the ingenuity of the designer, which becomes a limiting factor when dealing with complex devices. So far, there have only been limited attempts to merge the flexibility of optimization techniques with the well-understood physics of subwavelength structures: in  \cite{guo_realization_2023} a topology of conventional waveguides with interleaved subwavelength structures is proposed. By optimizing the shape of the waveguides, a variety of multiplexing devices with remarkable bandwidth were shown.

In terms of benchmarking, spot size converters that connect single-mode waveguides (typically $\thicksim$ 0.5\,µm wide) with grating couplers (with a typical width of 12\,µm) pose an interesting challenge for advanced design techniques [see \textbf{Figure \ref{fig:fig1}(a)}]. They require low insertion losses and broad bandwidth, yet when designed with traditional techniques exhibit large footprints to preserve adiabaticity or total internal reflection  \cite{xu_compact_2023,moreno-pozas_parabolic_2025}. \marktext{Lens-like structure have achieved excellent performance }\cite{yang_chip_2020}, with bandwidths exceeding 100\,nm, losses of $\thicksim$ 1\,dB and lengths ranging between 11\,µm and 14\,µm, with the best results reported for two etch-step fabrication processes \cite{luquegonzalez_ultracompact_2019,zhang_ultra-broadband_2021,wang_-chip_2019}. A variety of inverse-designed spot size converters have been demonstrated over the last years. A shape-optimized, 20\,µm-long homogenous taper was reported in \cite{zou_short_2014}, achieving sub-decibel insertion losses in a limited 45\,nm bandwidth. A 5\,µm-long transition based on pixel-like structures was demonstrated in \cite{liu_adiabatic_2019} but with a similarly limited bandwidth. A very interesting dual-lens configuration with 200\,nm bandwidth and multi-mode capabilities was reported in \cite{liu_achromatic_2024}, but required two etch-steps and at 26\,µm was comparatively long. Topology optimized structures have been shown to yield 8\,µm to 10\,µm-long devices, with reasonable insertion losses and bandwidths up to 100\,nm \cite{ma_topologically-optimized_2023}. A 5\,µm-long converter requiring two etch-steps with up to 100\,nm bandwidth was presented in \cite{wang_inverse-design_2024}, but exhibited comparatively high losses of 2\,dB. 

Here we propose a design approach that fully leverages both the inherent anisotropy of subwavelength metamaterials, and the flexibility afforded by inverse design to develop a high-performance spot size converter. Our topology, illustrated in Figure \ref{fig:fig1}(a), is composed of a fixed number of 25 cells that are optimized both in terms of their size and the equivalent anisotropic metamaterial they contain. This results in a comparatively small parameter space, which can be explored with fast 2D simulations. The equivalent metamaterials are then substituted by subwavelength structures which are locally refined using full 3D simulations. We experimentally demonstrate our approach with a single-etch step spot-size-converter that is only 7.2\,µm long yet exhibits measured sub-decibel losses in a 160\,nm bandwidth – to our knowledge this is the best combination of broad bandwidth and reduced footprint reported to date.\marktext{We furthermore show through simulation that ultra-low losses can be achieved with a more complex 81-cell configuration.}

\section{Design}

\begin{figure}
  \centering
  \includegraphics[width=0.5\linewidth]{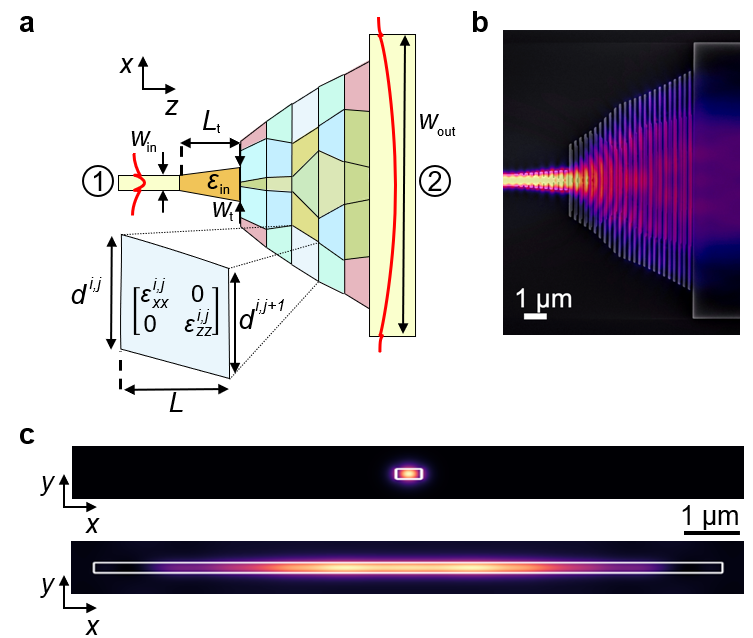}
  \caption{(a) Schematic of our topology for the spot-size converter, with each colored cell representing a different metamaterial. (b) Scanning electron microscope image of the fabricated spot-size converter; a full 3D-FDTD simulation of light propagation is superimposed. (c) Magnitude of the mode of the 500\,nm wide input waveguide (top) and simulated electric field in the 12\,µm wide output waveguide (bottom).}
  \label{fig:fig1}
\end{figure}

Referring to Figure \ref{fig:fig1}, we aim to design a spot size converter (SSC) that adapts the fundamental TE mode ($x$-polarized electric field) of a $W_{\mathrm{in}} =500$\,nm  wide waveguide to a $W_{\mathrm{out}}=12$\,µm wide output waveguide. For our design we consider a 220\,nm thick silicon layer, with a 2\,µm-thick buried $\mathrm{SiO_2}$ layer and cladding. The geometry we will optimize comprises a 5 × 5 matrix of SWG cells which are modeled with their equivalent anisotropic permittivity tensor $\overline{\overline{\varepsilon}}\,^{\mathrm{(i,j)}} = \mathrm{diag}[\varepsilon^\mathrm{(i,j)}_{xx},\varepsilon^\mathrm{(i,j)}_{zz}]$, where $\varepsilon_{xx}$ and $\varepsilon_{zz}$ are related via the properties of the subwavelength structure, as discussed below. The in-plane shape of the cells is controlled by the transversal distances $d^{\mathrm{(i,j)}}$ and a fixed length $L= 1$\,µm is used the direction of propagation. An additional transition of fixed length $L_\mathrm{t}=2.2$\,µm is placed between the input waveguide and the metamaterial matrix to reduce losses due to mode mismatch. Its permittivity $\varepsilon_{\mathrm{ini}}$  and final width $W_\mathrm{t}$ are also treated as optimization variable, allowing the taper to be considered as an additional cell in the problem. More detailed information about the geometry of the input taper is provided in the supporting information. Since the structure is symmetric with respect to $x=0$, only 3×6 transversal distances and 3×5 metamaterials need to be optimized.

To achieve broadband operation, we define the objective function as the insertion loss averaged over the wavelength range between $\lambda_1=1.5$\,µm and $\lambda_2=1.6$\,µm:
\begin{equation}
    f(\textbf{v})= \frac{1}{N}\sum_{i=1}^N {T}(\lambda_i,\textbf{v}),\, 
    {T} = 10  \log_{10} |S_{21}  |^2  
    \label{eq:eq1}
\end{equation}
Here \textbf{v} is a vector encompassing all the design variables and ${S_{21}}$ is the $S$-parameter characterizing the transmission from the fundamental TE mode of the narrow input (port 1) to the fundamental mode of the wide output (port 2), as shown in Figure \ref{fig:fig1}(a). To find a combination of parameters $\mathrm{\textbf{v}_{opt}}$ that minimizes $f$, i.e. 
\begin{equation}
    \textbf{v}_\mathrm{opt} = \mathrm{argmin_\textbf{v}} f(\mathrm{\textbf{v}}),
    \label{eq:eq2}
\end{equation}
the Covariance Matrix Adaption Evolution Strategy (CMA-ES) is employed \cite{hansen_reducing_2003}. The algorithm iterates through a loop where each iteration consists of a population of $p=10$ candidates sampled from a multi-variate normal distribution. Each candidate corresponds to a specific problem geometry $\mathbf{v}$, which is simulated electromagnetically to evaluate the objective function $f(\mathbf{v})$. Once all the candidates are evaluated, the algorithm calculates the next generation \cite{hansen_reducing_2003,hansen_cma_2016}.

\begin{figure}
  \centering
  \includegraphics[width=\linewidth]{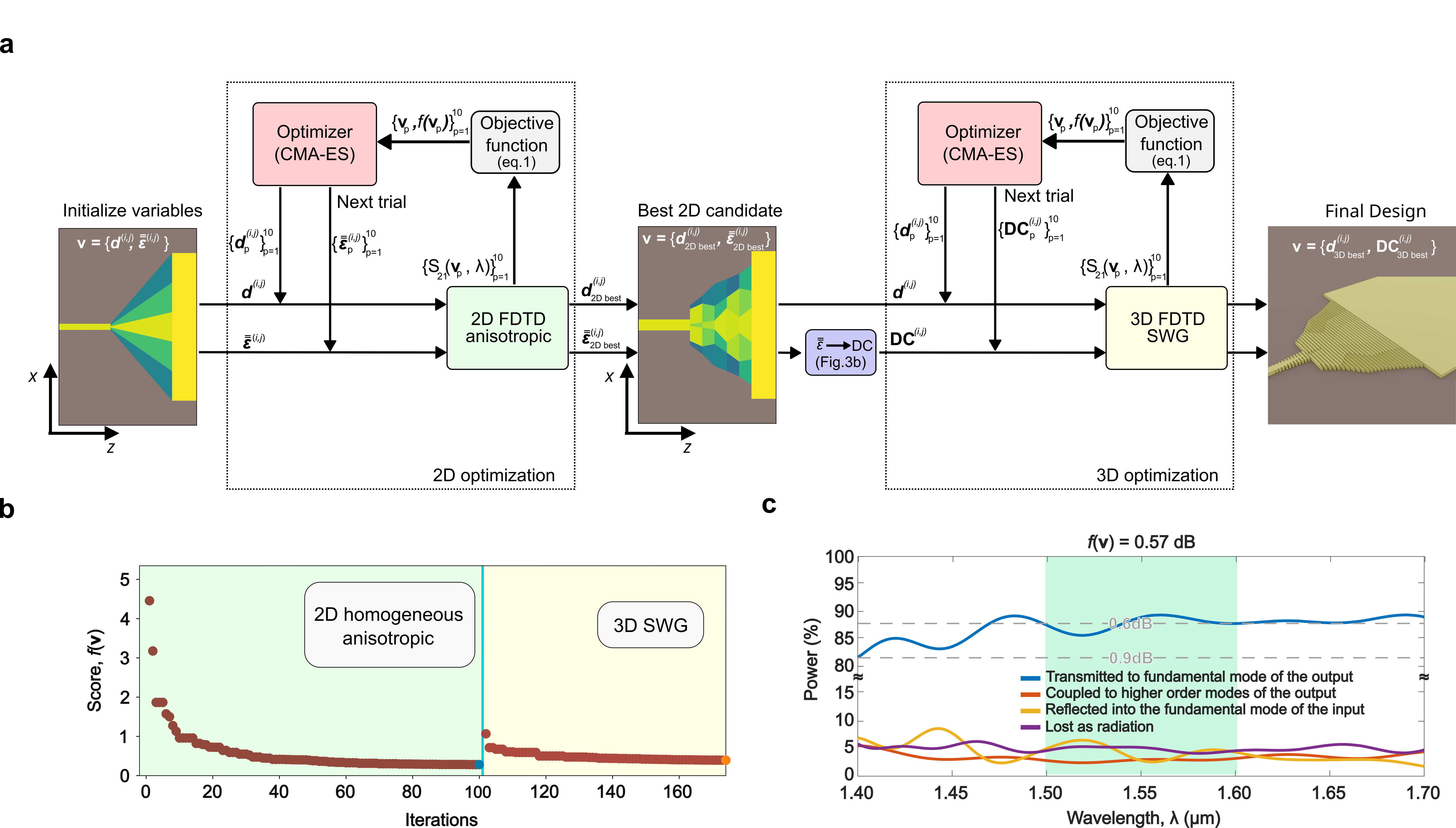}
  \caption{(a) Design methodology proposed in this work. The optimization process is divided in two stages: a first step with 2D simulations and homogenous anisotropic metamaterials enables extensive exploration of the design space, while a second step with 3D simulations of the full subwavelength structure is used to fine-tune the results. (b) Optimization history of the score function $f(\mathbf{v})$ for each iteration. \marktext{(c) 3D-FDTD simulation of the full sub-wavelength structured }$5\times5$\marktext{ cell device over a 300 nm bandwidth, showing both transmission to the fundamental output mode and the different sources of loss. The region shaded in green indicates the bandwidth in which the optimization is performed.}
  }
  \label{fig:fig2}
\end{figure}

\textbf{Figure \ref{fig:fig2}(a)} shows a schematic representation of the optimization process followed in this work. To enhance computational efficiency and reduce computation times, the optimization is initially carried out with 2D $(x, z)$ simulations, and the resulting structure is fed into the final 3D optimization. For the initial 2D optimization step, each SSC cell is modeled as a uniaxial crystal with a permittivity tensor, $\overline{\overline{\varepsilon}}\,^{\mathrm{(i,j)}}$, which fully characterizes the in-plane propagation \cite{halir_ultrabroadband_2016,perez-armenta_polarization-independent_2022}. The advantages of using this anisotropic model as opposed to a simpler isotropic one are discussed in the supporting information. The process to obtain the equivalent material model is illustrated in \textbf{Figure \ref{fig:fig4}(a)}. The SWG structure is analyzed with a Bloch-Floquet mode solver such as the MIT Photonic Bands (MPB) \cite{johnson_block-iterative_2001}. The structure extends indefinitely in the $x$ direction and is periodic in the propagation direction $z$. Solving for the in-plane polarized fundamental modes propagating in the $z$ and $x$ directions yields the components of the equivalent permittivity tensor $\overline{\overline{\varepsilon}}$, $\varepsilon_{xx}$ and $\varepsilon_{zz}$ respectively, which define the 2D material model of the SWG. To obtain different metamaterials properties we keep a constant pitch $\Lambda=200$\,nm (5 periods in each section of length $L=1$\,µm) while varying the duty-cycle of the SWG, defined as $a/\Lambda$  [see Figure \ref{fig:fig4}(a)]. Using this approach, a Look-Up Table (LUT) is created mapping each duty-cycle to its equivalent anisotropic metamaterial. To ensure that the resulting SWG is fabricable, the duty-cycle is limited between 30\% and 70\%, yielding a minimum feature size (MFS) of 60\,nm, which is compatible with electron-beam lithography (EBL). Since the top $\mathrm{SiO_2}$ cladding may not completely fill the trenches between the subwavelength stripes \cite{herrero-bermello_experimental_2020}, 60\,nm air-gaps were included in the model. \marktext{FIB-SEM inspection performed on the fabricated samples revealed triangular-shaped gaps, with areas comparable to those used in simulation. It has been shown that the exact shape of the gaps has no significant impact, as long as their overall area is well approximated }\cite{warshavsky_accurate_2025}. This results in the equivalent indices shown in Figure \ref{fig:fig4}(b), which constitute the look-up table to convert between the 2D homogeneous model and the full 3D structure. \marktext{We note from Figure} \ref{fig:fig4}\marktext{(b) that the strip-like SWG yields a strong anisotropy for the selected TE polarization, which is desirable to achieve an ultra-compact design. However, this strip-like SWG behaves quite differently for TM-polarization (see }\cite{luque-gonzalez_review_2021}\marktext{), which precludes polarization independent operation. Using a bricked metamaterial }\cite{perez-armenta_polarization-independent_2022}\marktext{, could enable polarization insensitivity while maintaining an analogous design methodology.}

Once the LUT has been created, the process outlined in Figure \ref{fig:fig2}(a) is followed. Referring to Figure \ref{fig:fig1}(a), the initial transition is initialized to $n_\mathrm{in}  = \sqrt{\varepsilon_\mathrm{in}}= 2.4$ and $W_\mathrm{t}=500$\,nm, while the metamaterial matrix is initialized in a GRIN-like manner, a strategy previously demonstrated to be effective for on-chip beam collimation [7]. Specifically, the refractive indices were set as follows: $n_\mathrm{xx}^\mathrm{(0,j)}  = 2.4$, $n_\mathrm{xx}^\mathrm{(1,j)}  = 2.2$ and $n_{xx}^\mathrm{(2,j)}  = 2.0$, with $n_{xx} = \sqrt{\varepsilon_{xx}}$. The corresponding $n_{zz} = \sqrt{\varepsilon_{zz}}$ values are obtained from Figure \ref{fig:fig4}(b). The transversal distances ${d^\mathrm{(i,j)}}$ are set so that the structure constitutes a linear taper from $W_\mathrm{in}= 500$\,nm to $W_\mathrm{out}  = 12$\,µm, i.e ${d^\mathrm{(i,j)}=[W_\mathrm{in}+(W_\mathrm{out}-W_\mathrm{in})j/5]/5}$. The input and output waveguide refractive index is $n = 2.85$, determined by applying the effective index method to a 220\,nm-thick SOI slab at $\lambda= 1550$\,nm  for TE polarization.

\begin{figure}
  \centering
  \includegraphics[width=0.5\linewidth]{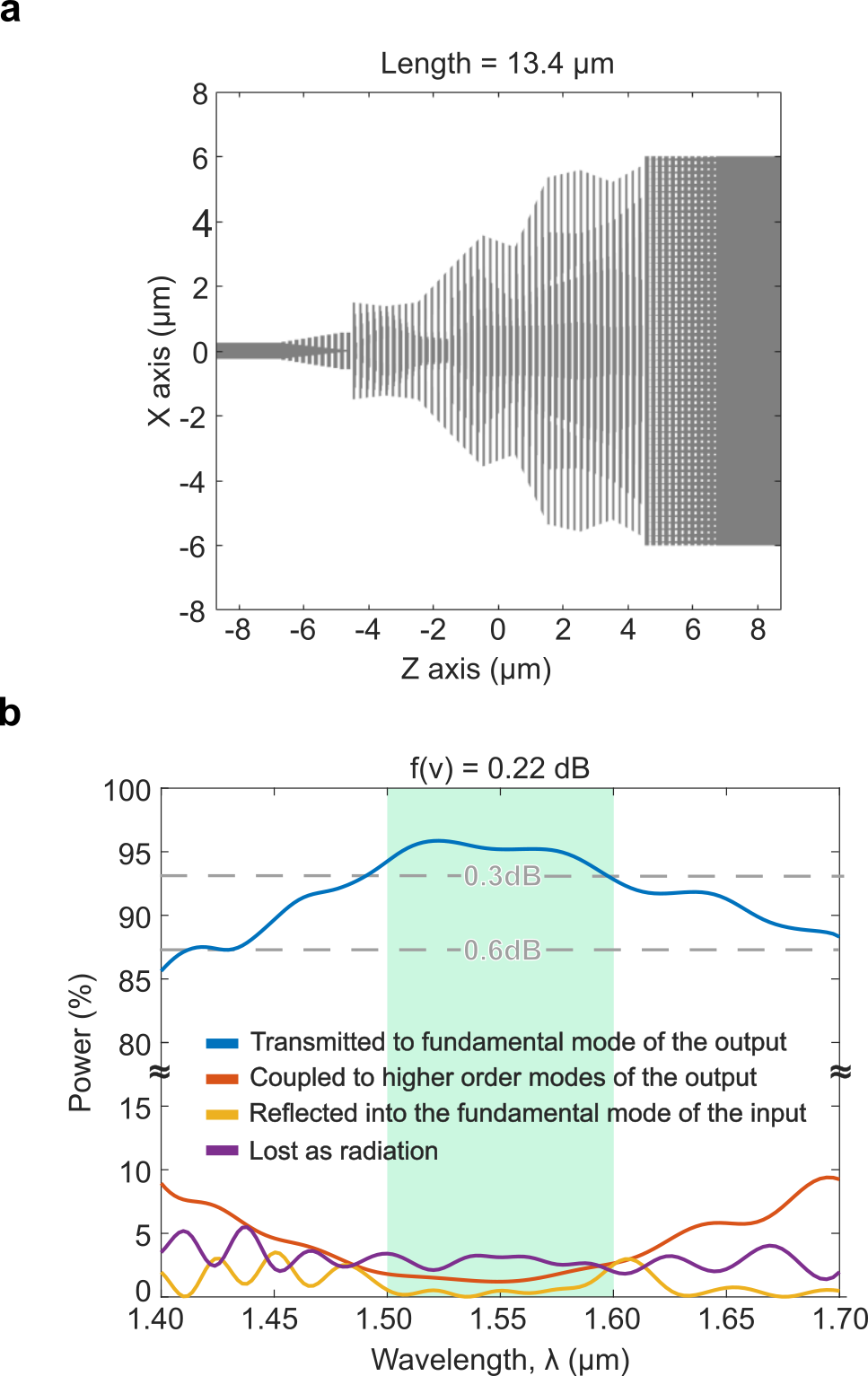}
  \caption{(a) Optimized spot-size-converter based on $9\times9$ cell configuration for ultra-low-loss. (b) 3D-FDTD simulation of transmission and reflection of the full sub-wavelength structured $9\times9$ cell device over a 300 nm bandwidth.}
  \label{fig:fig3}
\end{figure}

\begin{figure}
  \centering
  \includegraphics[width=0.5\linewidth]{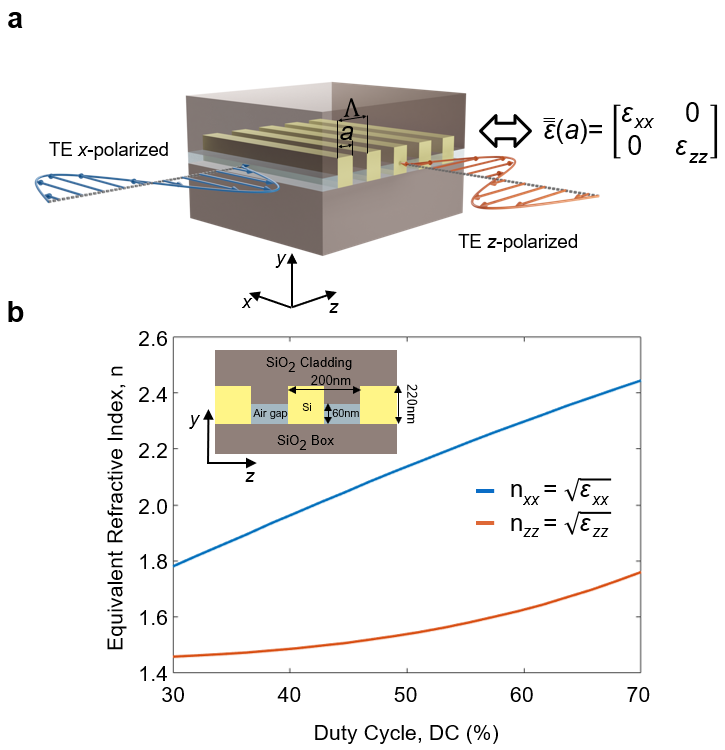}
  \caption{(a) Equivalent anisotropic 2D modeling of SWG metamaterials, including air gaps due to incomplete oxide filling. (b) Look-up table relating duty-cycle and equivalent indexes for a silicon-on-insulator SWG waveguide with period $\Lambda = 200$\,nm and thickness h = 220\,nm.  }
  \label{fig:fig4}
\end{figure}

The resulting 2D structure was simulated with the finite-difference time-domain (FDTD) simulator Meep to obtain the device $S$-parameters [see Figure \ref{fig:fig2}(a)]. Each iteration consists of $p=10$ electromagnetic simulations run in parallel. The $S_{21}$ parameter was used to compute the score function for each candidate in the iteration, as defined in \textbf{Equation \ref{eq:eq1}}. In the first iteration, the average insertion loss is $4.5\,\mathrm{dB}$, as shown in Figure \ref{fig:fig2}(b). The insertion loss decreases progressively during the optimization, reaching $0.2\,\mathrm{dB}$ after 100 2D iterations, beyond which no significant improvements are observed. Remarkably, substantial performance is achieved within the first 20 iterations, while subsequent iterations are devoted to fine-tuning the design for optimal results. The final iteration is considered the best 2D candidate (${d_\mathrm{2D\;best}^\mathrm{(i,j)}}$,$\overline{\overline{\varepsilon}}_\mathrm{2D\;best}^\mathrm{(i,j)}$) and serves as initialization for the optimization of the full 3D structure, using the LUT to map each homogeneous metamaterial cell into an SWG structure with the corresponding duty-cycle [see Figure \ref{fig:fig2}(a)]. Notably the first 3D simulation already yields an average insertion loss of only $1\,\mathrm{dB}$, which is reduced to $0.4\,\mathrm{dB}$ after 70 iterations – see Figure \ref{fig:fig2}(b). The dimensions of the final design are given in Tables S1 and Table S2 of the supporting information.Figure \ref{fig:fig2}\marktext{(c) shows full 3D FDTD simulations of the best design of the 3D optimization stage. Simulated losses are less than 0.9dB between 1400nm and 1700nm, and $\thicksim$ 0.6dB between 1500nm and 1700nm. The losses originate approximately equally from back-reflections, coupling to higher-order modes of the output waveguide, and radiation. The broad bandwidth of our spot-size-converter} is expected both from the compact size (7.2\,µm including the input taper), as well as the broadband behavior of subwavelength structures. \marktext{We note that the losses of the device could be further reduced to $\thicksim$ 0.2dB in the 1500nm to 1600nm band, using a larger $9\times9$ cell configuration, as discussed in section 4 of the supporting information.}   
The average simulation time for each 2D simulation was 1.5 minutes, whereas each 3D simulation required approximately 1.2 hours. This two-stage approach is therefore crucial for accelerating the convergence of the optimization process. The simulations were performed on the Picasso supercomputer \cite{scbi_available_nodate}, equipped with thousands of CPU cores and high-memory nodes, allowing the parallelization of tasks and significantly reducing computational time.\marktext{To demonstrate the flexibility and robustness of our approach, we also designed a more complex spot-size-converter, based on a }$9\times9$\marktext{ cell configuration, which exhibits losses of $\thicksim$ 0.2dB in the 1500nm to 1600nm band, as shown in Figure} \ref{fig:fig3}\marktext{. The details of this design a described in section 4 of the supporting information.}

\section{Experimental results}

\begin{figure}
  \centering
  \includegraphics[width=0.5\linewidth]{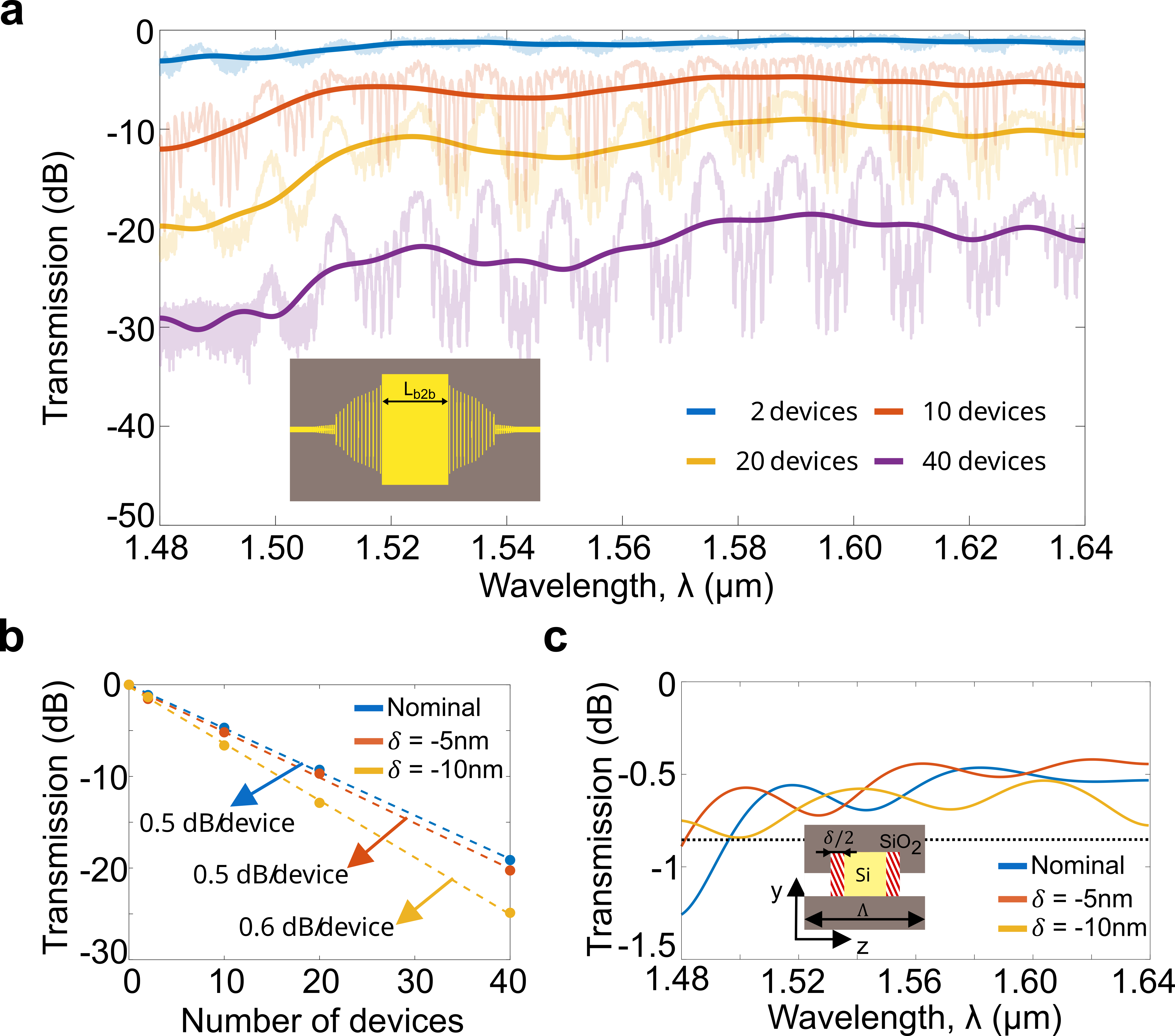}
  \caption{(a) Transmission measurements of different numbers of devices in back-to-back configuration. Pale lines represent raw data, while vivid lines indicate measurements processed with a minimum phase algorithm to eliminate reflections. (b) Transmitted power as a function of the number of devices with different biases at a wavelength of 1580\,nm. (c) Per-device transmission as function of wavelength for nominal design and two biases. For each wavelength the insertion loss is obtained from the slope of the linear regression performed on the cutback data.}
  \label{fig:fig5}
\end{figure}

To experimentally demonstrate \marktext{our design technique, the} $5\times5$\marktext{ cell device show in Figure} \ref{fig:fig2}\marktext{ was fabricated} on Applied Nanotools 220\,nm-thick SOI platform using electron beam lithography \cite{applied_nanotools_available_2024}. A bias in the duty-cycle was introduced in the different test structures as defined in the inset of \textbf{Figure \ref{fig:fig5}(c)} ($\delta= \pm5\,\mathrm{nm},$ $\pm10\,\mathrm{nm}$). In our measurements light from a tunable laser (Santec TSL-770) was coupled into the chip using a lensed fiber and a polarization controller, through an SWG edge coupler \cite{cheben_broadband_2015}. The polarization controllers along with a Glan-Thompson polarizer at the output, allowed precise control of the polarization state within the chip. At the chip output, light was focused onto a photodetector using a microscope objective. The transmission spectrum was obtained by sweeping the wavelength of the tunable laser and recording the corresponding output power. Figure \ref{fig:fig5}(a) presents the measurement data obtained by concatenating different numbers of devices for the nominal design, normalized to the transmission of a reference waveguide. The devices are arranged in a back-to-back configuration, as depicted in the inset, with a $L_\mathrm{b2b}=25\,$µm-long and 12\,µm-wide waveguide between devices and a 150\,µm-long single-mode waveguide between back-to-backs. These distances are chosen to ensure that the cavities formed by reflections from devices produce several oscillations within the 160\,nm measurement bandwidth. To eliminate these oscillations, a minimum phase technique \cite{halir_characterization_2009} has been applied in post-processing, which is shown as a solid line in Figure \ref{fig:fig5}(a). The figure shows measurements for up to 40 concatenated devices (20 back-to-back structures), showcasing the robustness and performance of the design. \marktext{We note that the ripple present in the raw measurement data is compatible with our simulated reflections of }$\thicksim$\marktext{ 4\% and its structure arises from the interplay of cavities that form in the test structures, as discussed in section 5 of the supporting material.} Figure \ref{fig:fig5}(b) depicts the linear regression at 1580\,nm for three flavors, with the slope representing the estimated device losses, which are below 0.5\,dB per device for the nominal design. Repeating this procedure across different wavelengths yields Figure \ref{fig:fig5}(c), which shows the transmission as a function of wavelength for various biases.  These results demonstrate that the devices can operate over a bandwidth of at least 160\,nm with insertion losses below $\thicksim$ 0.8\,dB. 

To compare this performance with the state-of-the-art summarized in Table 1, we introduce the following figure of merit which incorporates the measurement bandwidth, device length and expansion factor:
\begin{equation}
    \mathrm{FOM} = \frac{\mathrm{BW_{1dB}}}{\mathrm{device \: length}} \frac{W_\mathrm{out}}{W_\mathrm{in}} 
\end{equation}
A high FOM thus indicates a device that achieves a wide beam expansion over a broad bandwidth yet with a minimal length. As shown in Table 1, our spot-size converter improves on the state of the art by a significant margin, validating the potential of our metamaterial cell-based optimization approach. 

\begin{figure}
  \centering
  \includegraphics[width=0.5\linewidth]{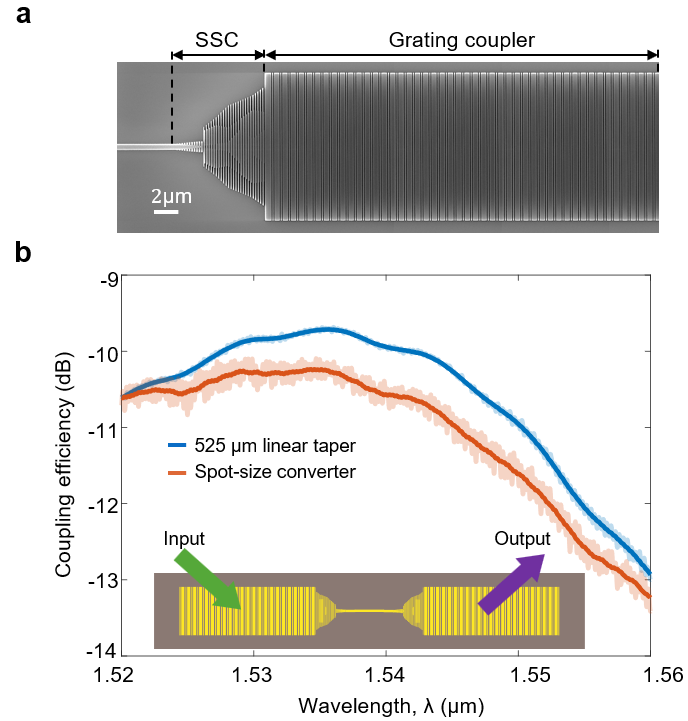}
  \caption{(a) Scanning electron microscopy image of a grating coupler feed by the spot-size converter designed in this work. (b) Measured transmission from the input grating coupler, through the spot-size-converter to a single-mode waveguide, and through a second spot-size-converter feeding the output grating coupler (orange line), compared to the same configuration with 525\,µm linear tapers (blue line) instead of the spot-size-converters. }
  \label{fig:fig6}
\end{figure}

As an additional verification, grating couplers were fed with our spot-size converter as shown in \textbf{Figure \ref{fig:fig6}(a)}. Identical gratings couplers fed by a 525\,µm linear tapers were also fabricated to serve as a reference, as the losses for this taper length are negligible. We measured transmission when coupling in-and-out through a pair of gratings as shown in the inset of Figure \ref{fig:fig6}(b), and calculated the coupling efficiency as the square root of this input-output transmission. As shown in Figure \ref{fig:fig6}(b) the nominal SSC design yields a measured coupling efficiency difference compared to the lossless linear taper of less than $0.7\,\mathrm{dB}$. This difference corresponds with the losses of the SSC and is consistent with the back-to-back measurements shown in Figure \ref{fig:fig5}(c). \marktext{There is a $\thicksim$ 0.2dB discrepancy between the measured losses and the losses predicted by simulations [Figure }\ref{fig:fig2}\marktext{(c)] which is attributed to minor fabrication imperfections.} While compare to the adiabatic taper a higher ripple due to reflections is observed [Figure \ref{fig:fig6}(b)], the overall length has been reduced by almost two orders of magnitude to 7.2\,µm. 

\section{Conclusion}

We have proposed a design methodology that leverages the optimization of cells of anisotropic subwavelength metamaterials to create a compact, high-performance spot-size converter. This design can expand a TE mode by a factor of 24 within a length of only 7.2 µm, with insertion loss below $\thicksim$ 0.8\,dB across a measured bandwidth of 160\,nm, thereby improving significantly on the state-of-the art. Indeed, our simulation predicts a potentially even larger bandwidth of 300\,nm. \marktext{This work serves as a proof of concept of the technique, demonstrating that it can produce broadband devices with a small number of cells and highly compact footprints, as well as ultra-low-loss devices composed of a larger number of cells. We believe that this will be essential for future applications requiring high integration density, such as programmable and quantum photonics.}

\textbf{Supporting Information} \par 
Supporting Information is available from the Wiley Online Library or from the author.
Data is available on Zenodo: https://doi.org/10.5281/zenodo.14640149.

\medskip
\textbf{Acknowledgements} \par 
We acknowledge funding through project PID2022-139540OB-I00 funded by MCIN/AEI\\/10.13039/501100011033. A. S. S. acknowledges an FPI scholarship with reference PRE2022-000085. The author thankfully acknowledges the computer resources technical expertise and assistance provided by the SCBI center of the University of Malaga.

\medskip

\break 

\clearpage 

\begin{table}[!htbp]
\begin{tabular}{cccccccccc}
\hline
\multicolumn{10}{c}{Table 1. Performance Comparison of SSC on SOI}                                                                                                                                            \\ \hline

Structure           & Ref      & $W_\mathrm{in}$[µm] & $W_\mathrm{out}$[µm] & $L_\mathrm{dev}$[µm] & $\mathrm{IL}_\mathrm{sim}$[nm] & $\mathrm{BW}_\mathrm{sim}$[nm] & $\mathrm{IL}_\mathrm{exp}$[dB] & $\mathrm{BW}_\mathrm{exp}$[nm] & $ \mathrm{FOM [\frac{nm}{\mathrm{\mu m}}]}$ \\ \hline
GRIN lens           & \cite{luquegonzalez_ultracompact_2019} & 0.5               & 15                 & 14                     & \textless 1         & \textgreater 350    & \textless 1         & \textgreater 130    & 279                   \\

Seg.taper           & \cite{zou_short_2014} & 0.5               & 12                 & 20                     & \textless 0.5       & \textgreater 60     & \textless 0.7       & \textgreater 45     & 54                    \\
Topo. Opt           & \cite{liu_adiabatic_2019} & 0.5               & 10                 & 5                      & \textless 1.5       & 40                  & \textless 1.5       & 40                  & 160                   \\
Luneburg lens\textsuperscript{1}       & \cite{zhang_ultra-broadband_2021} & 0.5               & 10                 & 11.2                   & \textless 1         & 740                 & \textless 1.5       & 220                 & 393                   \\
Metalens            & \cite{wang_-chip_2019} & 0.5               & 11                 & 13.7                   & \textless 0.55      & 200                 & \textless 0.8       & 100                 & 161                   \\
Metalens Opt.       & \cite{pascar_ultra-short_2024} & 0.5               & 15                 & 10                     & \textless 1         & 40                  & \textless 1         & 40                  & 120                   \\
Topo. Opt.          & \cite{ma_topologically-optimized_2023} & 0.5               & 10                 & 8                      & \textless 1         & 100                 & \textless 1         & 100                 & 250                   \\
Topo. Opt\textsuperscript{1}           & \cite{wang_inverse-design_2024} & 0.5               & 10                 & 5                      & \textless 1         & 100                 & \textless 2.2       & 100                 & 400                   \\
Parabolic reflector & \cite{xu_compact_2023} & 0.45              & 10                 & 32                     & \textless 0.2       & 200                 & \textless 0.15      & 100                 & 69                    \\
Achromatic lens\textsuperscript{1,2}     & \cite{liu_achromatic_2024} & 2                 & 10                 & 26.2                   & \textless 0.56      & 300                 & \textless 0.4       & 200                 & 38                    \\
This work           & -        & 0.5               & 12                 & 7.2                      & \textless 0.9         & 300                 & \textless 0.8         & 160                 & 533                   \\ \hline
\multicolumn{10}{l}{$ ^1$Two-step etching process.}                                                                                                                                                                 \\
\multicolumn{10}{l}{$ ^2$Multiples modes}                                                                                                                                                                           \\ \hline
\end{tabular}
\end{table}


\begin{figure}
\textbf{Table of Contents}\\
\medskip
  \includegraphics{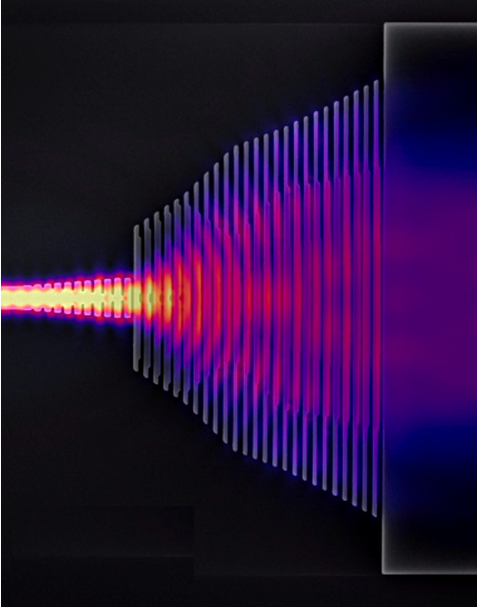}
  \medskip
  \caption*{Inverse design techniques are combined with subwavelength structures to design a compact, high-performance spot size converter. A cell-based topology is proposed, where each cell is composed of an equivalent homogenous metamaterial, resulting in a reduced parameter that speeds up optimization. A 7\,µm-long device that achieves a 24-fold beam expansion with sub-decibel losses over a 160\,nm bandwidth is experimentally demonstrated.}
\end{figure}

\clearpage



\section*{Supplementary material}

# Supporting Information

**Ultra-compact broadband spot size converter using metamaterial cell-based inverse design**

*Alejandro Sánchez-Sánchez\*, Carlos Pérez-Armenta, José Manuel Luque-González,*

*Alejandro Ortega-Moñux, J.Gonzalo Wangüemert-Pérez, Íñigo Molina-Fernández and*

*Robert Halir*

## 1. Final design data

Once the 3D optimization process is complete, the following tables provide the data required to define the device according to the cell definition in Fig. 1(a) of the main text. Specifically, table S1 presents the distances for each cell of the matrix, $d^{(i,j)}$, while Table S2 shows the duty cycle for each cell, $DC^{(i,j)}$. It is important to emphasize that the dimensions of the initial taper, which adapts the input waveguide to the SWG structure, are also included. Additionally, as discussed in the main document, the matrices are symmetrical with respect to the middle row.

Table S1. Distances matrix of the final design.

| $d^{(i,j)}$[µm] | Taper | | 0 | 1 | 2 | 3 | 4 | 5 |
|---|---|---|---|---|---|---|---|---|
| 2 | - | - | 0.306 | 0.630 | 1.800 | 1.500 | 1.203 | 2.079 |
| 1 | - | - | 1.234 | 1.520 | 1.188 | 0.842 | 2.342 | 1.413 |
| 0 | 0.5 | 0.884 | 0.128 | 0.449 | 0.719 | 2.805 | 1.260 | 2.772 |
| -1 | - | - | 1.234 | 1.520 | 1.188 | 0.842 | 2.342 | 1.413 |
| -2 | - | - | 0.306 | 0.630 | 1.800 | 1.500 | 1.203 | 2.079 |

Table S2. Duty cycle matrix of the final design.

| $DC^{(i,j)}$[%] | Taper ($DC_{in}$) | | 1 | 2 | 3 | 4 | 5 |
|---|---|---|---|---|---|---|---|
| 2 | - | - | 31.9 | 35.8 | 36.9 | 31.9 | 47.3 |
| 1 | - | - | 65.6 | 56.4 | 48.3 | 62.1 | 57.9 |
| 0 | 67.0 | 67.0 | 58.4 | 62.5 | 65.7 | 65.2 | 65.9 |
| -1 | - | - | 65.6 | 56.4 | 48.3 | 62.1 | 57.9 |
| -2 | - | - | 31.9 | 35.8 | 36.9 | 31.9 | 47.3 |

## 2. Input taper design

As discussed in the main text, achieving a high-performance device requires adapting the mode of the homogeneous waveguide to the SWG metamaterials used in the cell-based optimized device. The inverse design process determines the subwavelength duty cycle (DC$_{in}$) and width ($W_t$) that the taper must adapt to [see Fig. 1(a) in the manuscript]. Specifically, taper transitions from a homogenous silicon waveguide with a width of $W_{in} = 500$ nm to a SWG waveguide with a width $W_t = 884$ nm and a duty cycle of 67%, as obtained from the final 3D SWG optimization process (see tables S1 and S2)

The transition is achieved over 11 periods with $\Lambda = 200$ nm, maintaining a fixed cycle while varying the segment widths. The taper follows a double linear transition: the segment of length $a = DC \cdot \Lambda = 134$ nm increase linearly in width from 500 nm to 884 nm, while the segments of length $b = (1 - DC) \cdot \Lambda = 66$ nm decrease linearly in width from 500 nm to 0 nm, achieving the the desired SWG waveguide dimensions. The geometry of the taper can be seen in Figure S1(a) and Figure S1(b).

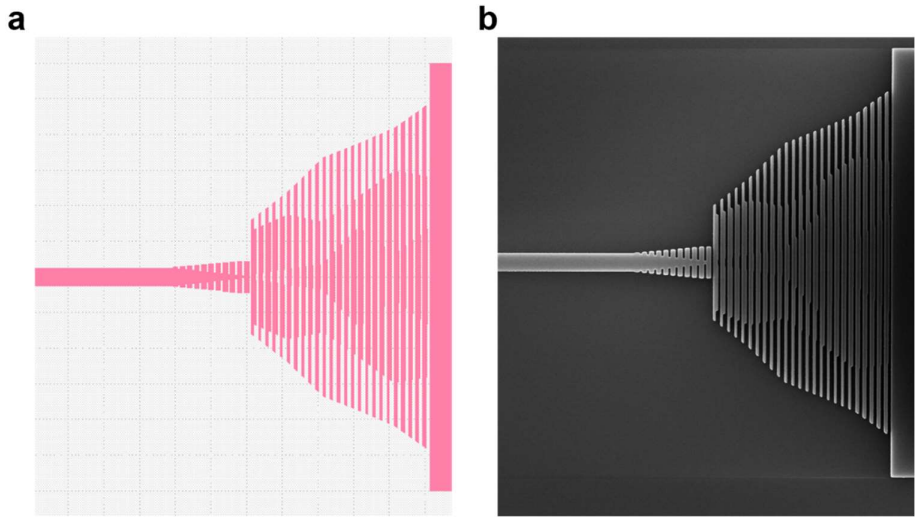

**a**　　　　　　　**b**

Figure S1. (a) Drawing of the final design included in the mask of fabrication. (b) Scanning electron microscope (SEM) image of the final design.

## 3. Advantage of modeling the anisotropy

In the device design, we model the anisotropy of the SWG to enhance the accuracy of the 2D homogeneous simulation. This section demonstrates the improvements achieved by introducing the anisotropic model. Figure S2(a) shows the optimization history of the score function, $f(\mathbf{v})$, comparing the isotropic model, $\bar{\bar{\varepsilon}}^{(i,j)} = \text{diag}[\varepsilon_{xx}^{(i,j)}, \varepsilon_{xx}^{(i,j)}]$, to the anisotropic model, $\bar{\bar{\varepsilon}}^{(i,j)} = \text{diag}[\varepsilon_{xx}^{(i,j)}, \varepsilon_{zz}^{(i,j)}]$, described in the main text. The starting points for both cases are equivalent, but the anisotropic model achieves faster improvements within just a few iterations. Moreover, the anisotropic model converges to lower losses. This behavior can be explained by the device's function as a lens. For lenses with the same maximum strength (determined by the contrast of the SWG that can be synthesized), the focal lengths is reduced when an anisotropic material is used, as demonstrated in [1].

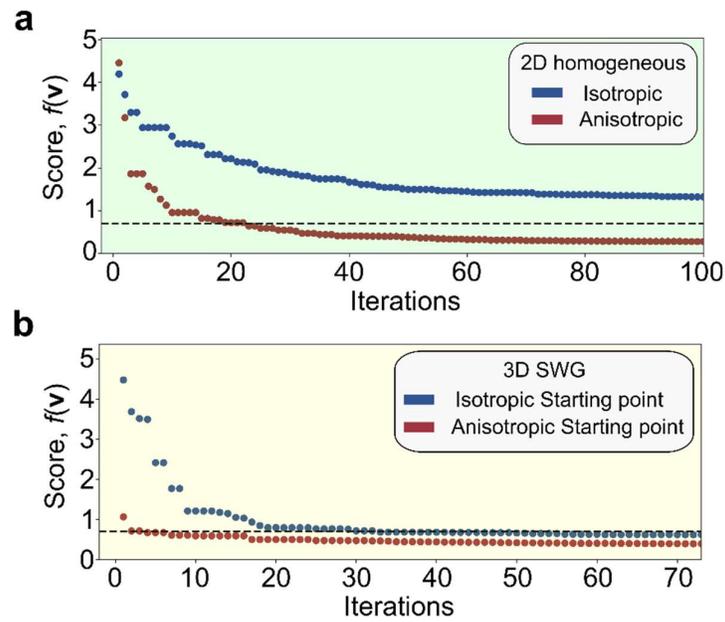

Figure S2. (a)-(b) Comparison between the optimization history of the score function, *f(v)* of the structure modeling the anisotropy versus one using an isotropic model for 2D simulation (a) and 3D simulation (b). Dashed line presents a score function of 0.7 dB in both cases.

Figure S2(b) illustrates the 3D optimization process of the devices, where the initialization corresponds to the best point obtained from Figure S2(a). It is important to note that the 3D simulation incorporates the SWG structures, where anisotropy is intrinsic. Consequently, when the best 2D candidate from the anisotropic model is used as the starting point, the initial performance is significantly better than when the starting point is provided by the isotropic model.

While the optimizer can still find a good solution starting from the best 2D candidate from the isotropic model, it requires more iterations and converges to a less optimal value. Hence, as the optimization process involves a degree of randomness and the 3D simulation are computationally expensive, incorporating the anisotropic model enhances the robustness of the optimizer, increasing the likelihood of identifying a local optimum efficiently.



\end{document}